\documentclass[12pt]{article}
\usepackage{amsmath}
\usepackage{amssymb}
\begin{document}
\vskip 2cm
\begin{center}
{\bf {\Large 
Nilpotent Charges of a Toy Model of Hodge Theory and an ${\cal N}$ = $2$ SUSY
Quantum Mechanical Model: (Anti-)Chiral Supervariable Approach }}

\vskip 1.5cm

{\sf T. Bhanja$^{(a)}${\footnote{On Leave of Absence from: Dept. of Physics, I. I. T. Guwahati, Guwahati - 781039 (Assam), India}}, 
 N. Srinivas$^{(a)}$ R. P. Malik$^{(a,b)}$}\\
$^{(a)}$ {\it Physics Department, Centre of Advanced Studies,}\\
{\it Banaras Hindu University, Varanasi - 221 005, (U.P.), India}\\

\vskip 0.1cm


\vskip 0.1cm

$^{(b)}$ {\it DST Centre for Interdisciplinary Mathematical Sciences,}\\
{\it Faculty of Science, Banaras Hindu University, Varanasi - 221 005, India}\\
{\small {\sf {E-mails: tapobroto.bhanja@gmail.com; seenunamani@gmail.com;
rpmalik1995@gmail.com}}}

\end{center}

\vskip 1.5cm

\noindent
{\bf Abstract:}
We derive the nilpotent (anti-)BRST and  (anti-)co-BRST symmetry transformations for 
the system of a toy model of Hodge theory (i.e. a rigid rotor) by exploiting the  (anti-)BRST and  
(anti-)co-BRST invariant restrictions on the (anti-)chiral supervariables that are defined on the 
appropriately chosen (1, 1)-dimensional super-submanifolds of the {\it general} (1, 2)-dimensional 
supermanifold on which our  system of a one (0 + 1)-dimensional (1D) toy model of Hodge theory is considered within the framework 
of the augmented version of the (anti-)chiral supervariable approach (ACSA) to Becchi-Rouet-Stora-Tyutin (BRST) formalism. The general (1, 2)-dimensional 
supermanifold is parameterized by the superspace coordinates ($t, \theta, \bar\theta$) where $t$ is 
the bosonic evolution parameter and ($\theta, \bar\theta$) are the  Grassmannian  variables which 
obey the standard fermionic relationships:  $ {\theta}^2 = {\bar\theta}^2 = 0, {\theta}\,{\bar\theta} + 
{\bar\theta}\,{\theta} = 0 $. We provide the geometrical interpretations for the  symmetry 
invariance and  nilpotency property. Furthermore, in our present endeavor, we establish the property of 
absolute anticommutativity of the conserved fermionic charges which is a completely {\it novel} and surprising observation 
in our present endeavor where we have considered {\it only} the (anti-)chiral supervariables. To corroborate the {\it novelty} of the
above observation, we apply this ACSA to an ${\cal N} = 2$ SUSY quantum mechanical (QM)  system of a free particle 
and show that the ${\cal N} = 2$ SUSY conserved and nilpotent charges  do {\it not} absolutely anticommute.

\vskip 1.2cm

\noindent
PACS numbers: 11.15.-q; 03.70.+k; 11.30.-j; 11.30.Pb; 11.30.Qc\\

\noindent
{\it Keywords:} {A 1D model of Hodge theory (i.e. a rigid rotor);  (anti-)BRST and (anti-)co-BRST symmetries;
(anti-)chiral supervariable approach; (anti-)BRST and (anti-)co-BRST invariant restrictions; conserved (anti-)BRST and 
(anti-)co-BRST charges; ${\cal N} = 2$ SUSY QM symmetries of a free particle; conserved and nilpotent SUSY charges;
nilpotency property;  absolute anticommutativity property}

\noindent
\section{Introduction}
The usual superfield approach [1-8] to Becchi-Rouet-Stora-Tyutin (BRST) formalism is one of the most
intuitive and instructive theoretical techniques that
provides the geometrical basis for the nilpotency as well as absolute anticommutativity properties
of the (anti-)BRST symmetries that are needed for the covariant canonical quantization of 
gauge theories which describe {\it three} (out of {\it four}) fundamental interactions of nature.
The above standard superfield approach leads to the derivation of (anti-)BRST symmetries 
{\it only} for the gauge and associated (anti-)ghost fields of a given
$D$-dimensional gauge theory (see, e.g. [3,4]).

The celebrated horizontality condition (HC) plays a key role in the standard superfield 
approach to BRST formalism where the Grassmannian components of the super curvature 2-form
(corresponding to a given  (non-)Abelian 1-form gauge theory) are set equal to zero so that the kinetic term
(i.e. a gauge invariant quantity) remains invariant in the sense that it does not
depend on the Grassmannian variables (see, e.g. [3,4] for details). This process of covariant reduction of 
the {\it super} curvature 2-form to the {\it ordinary} curvature 2-form (for a given (non-)Abelian 1-form
gauge theory) leads to the derivation of (anti-)BRST symmetry transformations for the gauge 
and associated (anti-)ghost fields of the above mentioned (non-)Abelian 1-form gauge theory
{\it without} any interaction with matter fields [3,4].

The above standard superfield approach [1-8]  has been now extended systematically 
(see, e.g. [9-12]) so that one could derive the (anti-)BRST symmetry transformations for the  {\it matter}, gauge and 
(anti-)ghost fields {\it{together}}  for an {\it interacting} gauge theory. The extended version 
(see, e.g. [9-12]) of the standard superfield formalism [1-8] has been christened as the 
augmented version of superfield  approach (AVSA). In the above superfield formalisms [1-12], the ordinary fields and spacetime 
coordinates of a given $D$-dimensional gauge theory are generalized to the corresponding superfields and superspace coordinates (which
characterize the appropriately chosen$(D, 2)$-dimensional supermanifold) and the {\it full} super expansion of the
superfields are taken into account (on the above supermanifold).

The central theme of our present investigation is to apply the (anti-)chiral 
supervariable approach (ACSA) to  derive the (anti-)BRST as well as the (anti-)co-BRST symmetry 
transformation for the one (0 + 1)-dimensional (1D) system of a rigid rotor which has been proven to be a model
for the Hodge theory in [13]. One of the key features of the ACSA is the observation that we
consider {\it only} the (anti-)chiral super expansions of the supervariables which are defined on the (1, 1)-dimensional 
super-submanifolds of the general (1, 2)-dimensional supermanifold on which our 1D theory is generalized. We apply the
physically motivated restrictions on the (anti-)chiral supervariables of the theory to obtain the (anti-)BRST and
(anti-)co-BRST symmetries of the 1D toy model of Hodge theory.

We have christened our present approach as the (anti-)chiral {\it supervariable} approach 
to BRST formalism because, first of all, we use only (anti-)chiral supervariables
for our discussion which have (anti-)chiral super expansions. Furthermore, we observe that the limiting case 
(when the Grassmannian variables are set equal to zero), we obtain a {\it variable}
from the supervariable (cf. Eqn. (7) below) which is a function of ``time'' only. 
On the contrary, in the standard  superfield formalism [1-8], we obtain a {\it field} which
 is a function of {\it spacetime} in the limiting case  when the Grassmannian variables are set equal
to zero in a super expansion of the superfield (within the framework of AVSA to BRST formalism).

One  of  the novel observations of our present endeavor is the emergence of the absolute
anticommutativity of the (anti-)BRST and (anti-)co-BRST  {\it charges} even though we consider only the
(anti-)chiral supervariables that are defined on the (1, 1)-dimensional (anti-)chiral 
super-submanifolds of the {\it general} (1, 2)-dimensional supermanifold. The other key result of our present investigation
is the observation that we do {\it not} use the (dual-)horizontality condition {\it anywhere}.
Instead, we exploit the beauty and strength of the (anti-)BRST and (anti-)co-BRST invariant
restrictions on the (anti-)chiral supervariables to derive the 
nilpotent (anti-)BRST and (anti-)co-BRST symmetry transformations.

Against the backdrop of the above arguments and claims, it is pertinent to point out that the
symmetry invariance has played a very decisive role in our earlier works [14-17] on the ACSA
 to ${\mathcal N} = 2$ supersymmetric (SUSY) quantum mechanical (QM) models where we have been 
able to derive the appropriate nilpotent ${\mathcal N} = 2$ SUSY transformations by exploiting
the SUSY invariant restrictions on the (anti-)chiral supervariables that are defined on the
(1, 1)-dimensional super-submanifolds of the {\it general} (1, 2)-dimensional supermanifold on 
which the usual SUSY QM theory is generalized. We have been able to capture the nilpotency of the 
${\mathcal N} = 2$ SUSY conserved and nilpotent charges. However, these charges have been found
{\it not} to respect the absolute anticommutativity property within the framework of ACSA to 
${\mathcal N} = 2$ SUSY QM models (see, e.g., Appendix B). Needless to say, the symmetry invariant
restrictions have reach and range that encompass in their folds 
the derivation of {\it nilpotent} symmetry transformations for the ${\mathcal N} = 2$ quantum
mechanical models as well as the (anti-) BRST invariant $D$-dimensional gauge theories.

The following factors have spurred our curiosity in pursuing our present investigation.
First, the (dual-)horizontality conditions are {\it mathematical} in nature. Thus, it 
is essential for us to provide an alternative to them by some physically motivated 
restrictions. We have accomplished this goal in our present investigation. Second, we have already 
applied our present theoretical technique 
in the context of the 4D Abelian 2-form gauge theory [18] and 2D self-dual
bosonic field theory [19] to derive the (anti-)BRST and (anti-)co-BRST 
symmetry transformations. Thus, it is urgent for us to look for the validity of this 
method in the context of some other systems. We have achieved this objective in our present 
investigation. Third, we have proven the absolute anticommutativity of the (anti-)BRST and 
(anti-)co-BRST charges {\it despite} the fact that we have considered only the (anti-)chiral 
supervariables in our present investigation. This is a completely {\it novel} observation in our present 
endeavor {\it vis-{\`a}-vis} the application of ACSA to ${\mathcal N} = 2$ SUSY QM models where absolute 
anticommutativity property does {\it not} exist.
Finally, our present and earlier works [18, 19] are our modest steps towards providing 
variety and theoretical richness in the context of superfield/supervariable approach to {\it quantum} gauge 
theories of all varieties.

Our present paper is organized as follows. In Sec. 2, we recapitulate the bare essentials of the nilpotent
(anti-)BRST and (anti-)co-BRST symmetry transformations of the Lagrangian for the toy model of a rigid rotor.
Our Sec. 3 deals with the derivation of (anti-)BRST transformations by exploiting the (anti-)BRST invariant
restrictions on the appropriate (anti-)chiral supervariables. Sec. 4 is devoted to the precise determination of  
(anti-)co-BRST symmetry transformations from the appropriate (anti-)co-BRST invariant restrictions on the (anti-)chiral 
supervariables. In Sec. 5, we capture the symmetry invariance of the Lagrangian
for our toy model of Hodge theory (i.e. a rigid rotor) within the framework of ACSA. Our Sec. 6 deals with the proof of
absolute anticommutativity and nilpotency 
of the conserved charges within the framework of augmented supervariable formalism (i.e. ACSA to BRST formalism). Finally,
we summarize our key results and point out a few future directions for further investigations in our
Sec. 7.

In our Appendix A, we comment on the {\it natural} emergence of absolute anticommutativity of the (anti-)BRST 
and (anti-)co-BRST charges as well as the  (anti-)BRST and (anti-)co-BRST symmetry 
transformations within the framework of superfield approach to BRST formalism where the {\it full}
expansions of the superfields, unlike our present endeavor with the (anti-)chiral supervariables, have been taken into account.
To explain the novelty of our observation (within the framework of ACSA), we discuss, in our Appendix B, a 
simple case of ${\mathcal N} = 2$ SUSY {\it free particle} and show that the application of ACSA to this 
system does {\it not} lead to the anticommutativity of the conserved and nilpotent 
${\mathcal N} = 2$ SUSY charges.

\section{Preliminaries: Nilpotent Symmetries}

Let us start off with the following (anti-)BRST and (anti-)co-BRST invariant first-order Lagrangian 
for the toy model of a Hodge theory (i.e. a rigid rotor 
with the mass parameter $m = 1$) as (see, e.g. [20,13] for details)
\begin{eqnarray}
L_b = \dot r \, p_r + \dot\vartheta \, p_{\vartheta} - \frac{p_{\vartheta}^2}{2r^2} 
- \lambda \,(r - a) + B\,(\dot \lambda - p_r)
+ \frac{B^2}{2} - i\, \dot{\bar C}\,\dot C + i\, \bar C\, C,
\end{eqnarray}
where ($r, \vartheta$) are the polar coordinates, ($\dot r, \dot\vartheta$) are the generalized
velocities and ($p_r, p_\vartheta$) are their corresponding conjugate momenta, $\lambda$ is a  
``gauge'' variable, $B$ is the Nakanishi-Lautrup type of auxiliary variable and $(\bar C)\,C$ 
(with $C^2 = {\bar C}^2 = 0,\,\, C\,\bar C + \bar C\,C = 0$) are the (anti-)ghost fermionic 
variables. All these variables are function of the evolution ``time" parameter {\it t}
and an overdot on any arbitrary variable denotes a derivative w.r.t. it (i.e. $\dot C = \frac{dC}{dt}$, etc.)

The action integral $S = \int dt\, L_{b}$ respects the following infinitesimal,   continuous and nilpotent 
($s_{(a)b}^2 = 0$) (anti-)BRST symmetry transformations ($s_{(a)b}$) (see, e.g. [20])
\begin{eqnarray}
&& s_{ab}\, \lambda = \dot {\bar C}, \qquad  s_{ab}\, C = - i B, \qquad  s_{ab}\, p_r = - \bar C,
\;\;\quad  s_{ab}\, [\bar C, r, \vartheta, p_{\vartheta}, B] = 0,\nonumber\\
&& s_b\, \lambda = \dot C, \qquad \,\,s_b \, \bar C = i B, \quad\qquad \, s_b\, p_r = - C,\,\, 
\qquad s_b \,[C, r, \vartheta, p_{\vartheta}, B] = 0,
\end{eqnarray}
because the Lagrangian $L_b$ transforms to the total time derivatives:
\begin{eqnarray}
s_{ab}\, L_b = \frac{d}{dt}\bigl [B\,\dot {\bar C} - \bar C\,(r-a)\bigr],\qquad\quad
s_b\, L_b = \frac{d}{dt}\bigl [B\,\dot C - C\,(r-a)\bigr].
\end{eqnarray}
There is another set of nilpotent ($s_{(a)d}^2 = 0$) continuous symmetry transformations ($s_{(a)d}$) 
in our theory [13]. These (anti-)dual [or (anti-)co] BRST symmetries ($s_{(a)d}$)
\begin{eqnarray}
&& s_d\, \lambda = \bar C, \qquad\,\,  s_d\, C = i\, (r-a), \,\quad\qquad  
s_d\, p_r = \dot{\bar C},\qquad\,\,  s_d \,[\bar C, r, \vartheta, p_{\vartheta}, B] = 0,\nonumber\\
&& s_{ad}\, \lambda = C, \qquad s_{ad}\, \bar C = - i\, (r - a), 
\qquad  s_{ad}\, p_r =  \dot C,\qquad  s_{ad}\, [ C, r, \vartheta, p_{\vartheta}, B] = 0,
\end{eqnarray}
leave the Lagrangian {\it absolutely} invariant (i.e. $s_{(a)d}\, L_b = 0$).

The above continuous symmetry transformations lead to the derivation of conserved charges due to
Noether's theorem. It can be readily checked, using the standard techniques of Noether's theorem, 
that the following conserved charges
\begin{eqnarray}
&& Q_b = B\,\dot C - \dot B\, C, \qquad\quad Q_{ab} = B\,\dot {\bar C} - \dot B\,\bar C,\nonumber\\
&& Q_d = B\,\bar C + \dot B\, \dot{\bar C}, \qquad\quad Q_{ad} = B\, C + \dot B\,\dot C,
\end{eqnarray}
are the generators of the above nilpotent symmetry transformations because of the
fact that the following is true, namely;
\begin{eqnarray}
s_r \, \omega(t) = \pm\, i\, \bigl [\omega(t), Q_r \bigr ]_{\pm}, \qquad\qquad r = b, \,ab, \,d, \,ad,
\end{eqnarray}
where  $\omega (= r, \vartheta, p_r, p_{\vartheta}, \lambda, C, \bar C, B)$ is the 
generic variable of our theory and  the $(\pm)$ signs (as the subscripts on the 
square bracket) correspond to the (anti)commutators depending on 
the (fermionic)bosonic nature of the 
generic variable $\omega(t)$. It turns out that the above conserved charges are {\it also}
nilpotent $(Q_{r}^{2} = 0, r = b, ab, d, ad)$ of  order two.

We end this section with the following remarks. First, the (anti-)BRST transformations
leave the kinetic term  $[ (\dot{\vartheta}\,p_{\vartheta}) - ({p^2_{\vartheta}}/{2r^2}) 
= \frac{1}{2}\,{r}^2\,{\dot{\vartheta}}^2 \, \equiv\,\frac{1}{2} v^2] $ 
 invariant (where $v$ is the linear velocity) (see, e.g. [13] for details). Second, the gauge-fixing term 
$ (\dot{\lambda} - p_r) $ remains unchanged
 under the (anti-)co-BRST symmetry transformations. Third, there is a {\it unique } 
 bosonic symmetry in the theory which is obtained by the appropriate anticommutators between 
 the (anti-)BRST and (anti-)co-BRST symmetry transformations. Fourth, there exists a ghost-scale
 symmetry in the theory, too. Fifth, the kinetic and gauge-fixing terms owe their origin 
 to the nilpotent exterior and co-exterior derivatives of differential geometry (see, e.g. [13] for details).
 Sixth, we note that the physical constraint $(r - a) = 0$ remains invariant  [i.e. $s_{(a)b}\, (r - a) = 0$]
 under the (anti-)BRST as well as (anti-)co-BRST symmetry transformations which ought to be true for a model of Hodge theory.
 Finally, we have established, in our earlier work [13], that our present 
 toy model is an example for the Hodge theory where the symmetries of the theory
 provide  physical realizations of the de Rham cohomological operators of 
 differential geometry (see, e.g. [21-24]).

\noindent
\section{(Anti-)BRST Symmetries: Supervariable Approach}

To derive the (anti-)BRST symmetries (2) within the framework of supervariable approach, 
first of all, we generalize the dynamical variables (e.g. $\lambda,\, C,\, \bar C,\, p_r$)
of the Lagrangian (1) onto the (1, 1)-dimensional {\it anti-chiral} super-submanifold 
 as follows
\begin{eqnarray}
&& \lambda(t) \rightarrow {\Lambda}(t, \bar{\theta}) = \lambda (t)  + \bar{\theta}\, R(t),
\qquad\qquad C(t) \rightarrow F(t,  \bar{\theta}) = C(t)  + i\,\bar{\theta} \, B_1(t),\nonumber\\
&& \bar{C}(t) \rightarrow \bar{F}(t, \bar{\theta}) = \bar{C}(t) + i\,\bar{\theta}\,B_2(t),
\quad\quad\,\,\,\, p_r(t)\rightarrow P_r(t,\bar{\theta}) = p_r(t) + \bar{\theta} \,K(t),
\end{eqnarray}
where the expansions on the r.h.s. contain the secondary variables $(R(t), B_1(t), B_2(t), K(t))$ which are 
function of {\it only} the evolution parameter $ t$ and, ultimately, they  have to be 
determined in terms of the basic and auxiliary variables of the theory (cf. Eq. (1)). 
We note that, in the limit $ \bar{\theta} = 0 $, we retrieve back our usual variables 
($\lambda (t), C(t), \bar C(t), p_r(t) $). We further observe that the anti-chiral 
super-submanifold is characterized by ($t, \bar\theta$). As a consequence, {\it all} the 
supervariables, defined on this super-submanifold, are function of ($t, \bar\theta$).

We mention, in passing, that we have {\it not} taken the expansions for the supervariables [corresponding
to the ordinary variables $(r, \vartheta, p_{\vartheta},B)$] because these are (anti-)BRST invariant
 (i.e. $s_{(a)b}[r, \vartheta, p_{\vartheta}, B] = 0$). As a consequence, we have the following restrictions:
\begin{eqnarray}
&& r(t)\rightarrow \tilde R^{(\cal B)}(t, \bar\theta) = r(t),\qquad 
\qquad \,\,p_\vartheta \rightarrow P^{(\cal B)}_\vartheta (t,\bar\theta) = p_{\vartheta}(t), \nonumber\\
&& \vartheta(t)  \rightarrow \Theta^{(\cal B)}(t,\bar\theta) = \vartheta(t),  
\qquad \qquad B(t) \rightarrow \Tilde B^{(\cal B)}(t,\bar\theta) = B(t).
\end{eqnarray}
The above equations demonstrate that we have {\it no} (anti-)BRST symmetry transformations 
for the variables $[r(t), \vartheta(t), p_{\vartheta}(t),B(t)]$ as would become clear later. 
The superscript ($\cal B$) on the supervariables denote that these have been obtained after 
the BRST invariant restrictions that have been imposed on them. These restrictions state 
that such kind of ``physical'' quantities should {\it not} depend on ``soul" coordinate 
$\bar\theta$ of the anti-chiral supervariables [25]  within the framework of augmented 
(anti-)chiral supervariable approach.

To determine the secondary variables of (7) (in terms of the basic and auxiliary variables of
(1)), we shall invoke the basic idea of augmented supervariable formalism [9-12]  which
requires, first of all, to determine the appropriate BRST invariant quantities. For instance, 
it can be readily checked that the following is true, namely;
\begin{eqnarray}
&&s_b (r, \vartheta, p_{\vartheta}, B, C) = 0,\quad \, s_b (\lambda\,\dot C) = 0, \quad\quad s_b (p_r\, C) = 0, \nonumber\\
&& s_b(\dot B\, \lambda + i\,\dot{\bar C}\, \dot C) = 0, \quad \,\,
s_b(\dot p_r + \lambda) = 0, \quad s_b(B\, p_r - i\,{\bar C}\, C) = 0.
\end{eqnarray}  
The second step is to generalize these quantities (present in the parenthesis) onto the anti-chiral
super-submanifold and demand $\bar\theta$-independence of these quantities. Such requirement is essential because 
the ``soul" coordinate $\bar\theta$ is not a ``physical" object whereas the BRST-invariant
quantities are [25]. Thus, we have the following non-trivial restrictions  (besides (8)) on the 
combinations of (super)variables, namely;
\begin{eqnarray}
&& F^{(\cal B)} (t, \bar\theta) = C(t), \quad P_r (t,\bar\theta) \,F^{(\cal B)} (t, \bar\theta) = p_r (t)\, C(t), 
\quad {\dot P}_r (t, \bar\theta) + \Lambda(t, \bar\theta) = {\dot p}_r (t) + \lambda (t),\nonumber\\
&& \dot{\tilde B}^{(\cal B)} (t,\bar\theta)\, \Lambda (t,\bar\theta) 
+ i\, \dot {\bar F} (t, \bar\theta)\, {\dot F}^{(\cal B)} (t, \bar\theta) 
= \dot B (t)\, \lambda (t) + i\,\dot{\bar C}\, \dot C,\nonumber\\
&& {\tilde B}^{(\cal B)}(t, \bar\theta) \,P_r(t, \bar\theta) 
- i\, {\bar F} (t, \bar\theta)\, F^{(\cal B)} (t, \bar\theta) = B(t)\, p_r(t) - i\, {\bar C}(t) \, C(t). 
\end{eqnarray}
The substitution of expansions in (7) and (8), into the above, leads to the following {\it useful} relationships
amongst the secondary and basic variables, namely;
\begin{eqnarray}
&& R(t)\, \dot C(t) = 0, \quad\qquad K(t)\,C(t) = 0, \quad\qquad  {\dot K(t)}  + R(t) = 0, \nonumber\\
&& \dot B(t)\, R(t) - {\dot B}_2 (t)\, {\dot C}(t) = 0, \quad\qquad B(t)\, K(t) + {B_2}(t) \, C(t) = 0. 
\end{eqnarray}
It is straightforward to note that the following solutions
\begin{eqnarray}
R(t) = \dot C(t), \quad\qquad {B}_2 (t) = B(t), \quad\qquad K(t) = - \,C(t),\qquad B_1(t) = 0,
\end{eqnarray}
satisfy {\it all} the relationships in (11). These solutions can be modified modulo a constant 
overall factor. This freedom is always present in our original transformations  (2), too. 
In fact, we have made one of the simplest possible choices here. Ultimately, we have the 
following expansions  for the supervariables ({\it vis-{\`a}-vis} the BRST transformations in 
(2)): 
\begin{eqnarray}
&&\Lambda^{(\cal B)}(t, \bar{\theta})\, =\, \lambda(t) \,+ \,\bar{\theta}\,(\dot{C}) \,\equiv  \,\lambda (t)\,+ 
\,\bar{\theta}\, (s_{b} \, \lambda(t)), \nonumber\\
&&F^{(\cal B)}(t, \bar{\theta})\, =\, C(t) \,+ \,\bar{\theta}\,(0) \,\equiv \, C(t) \, 
+ \, \bar{\theta}\, (s_{b}\,C(t)),\nonumber\\
&&{\bar{F}}^{(\cal B)}(t,  \bar{\theta})\, = \,\bar{C}(t)\, +\, \bar{\theta}\, (i B(t))\, \equiv\,  \bar{C}(t) \, 
+\, \bar{\theta}\, (s_{b}\,\bar{C}(t)),\nonumber\\
&&{P_r}^{(\cal B)}(t, \bar{\theta})\, =\,  p_r (t)\, + \, \bar{\theta}\, (-\, C(t))\, 
\equiv \, p_r (t) \, + \,\bar{\theta}\, (s_{b}\,p_r (t)).
\end{eqnarray}
A close observation of (13) demonstrates that we have already obtained the non-trivial BRST symmetry 
transformations $s_b$ of our theory (cf. (2)). The trivial BRST symmetry transformations 
$s_{b}[r, \vartheta, p_{\vartheta}, B] = 0$ have already been captured in (8) if we take 
the analogy with (13) and interpret the coefficient of $\bar\theta$ as the BRST transformations ($s_b$).

The above expansions in (13) lead to the following equalities/mappings
\begin{eqnarray}
\frac{\partial}{\partial{\bar\theta}}\,\Omega^{(\cal B)}(t, \bar{\theta}) =
\, s_{b}\,\omega(t)\, \equiv \, \pm\, i\, \bigl[\omega(t), \, Q_b \bigr]_{\pm},                   
\qquad \qquad s_b\, \leftrightarrow  \, \partial_{\bar\theta} \, \leftrightarrow  \, Q_b,
\end{eqnarray}
where $ \Omega^{(\cal B)}(t, \bar{\theta}) $ is the generic supervariable  derived after the application 
of the BRST invariant restrictions (8) and (10) and $ \omega(t) $ is the generic ordinary variable 
of the starting Lagrangian [quoted in (1)]. In the above equation (14), we have taken the general
 definition of a generator  of a symmetry transformation [as given in (6)]. Thus, we note that the 
 operators ($\partial_{\bar\theta}, \, s_b,\, Q_b $) are inter-related. In other words, the nilpotency
 of $ \partial_{\bar\theta},  \, s_b$ and $Q_b$ are deeply intertwined. The relationship (14) provides 
 the geometrical interpretation for $s_b$ in the language of translational generator $\partial_{\bar\theta}$ 
 along the Grassmannian direction $\bar\theta$ of the anti-chiral (1, 1)-dimensional 
super-submanifold (of the {\it general} supermanifold).

Now we concentrate on the derivation of anti-BRST symmetry transformation $ s_{ab}$ of (2) by 
invoking the virtues of anti-BRST invariant restrictions on the {\it chiral} supervariables that are 
defined on the (1, 1)-dimensional {\it chiral} super-submanifold. First of all, we have the following
generalizations of the dynamical variables ($\lambda, C, \bar C, p_r$) onto the chiral (1, 1)-dimensional
 super-submanifold, namely;
\begin{eqnarray} 
&& \lambda(t)\,\rightarrow\, \Lambda(t, \theta)\, = \,\lambda(t)\, + \,\theta\, \bar{R}(t), 
\qquad\quad\,\, C(t)\, \rightarrow \,F(t, \theta)\, = \,C(t)\, +\, i \, \theta \, \bar{B}_{1}(t), \nonumber\\
&& \bar{C}(t) \, \rightarrow  \,\bar{F}(t, \theta) \, =  \,\bar{C}(t) \, + \, i \, \theta \, \bar{B}_2(t),  
\qquad p_r(t) \, \rightarrow \, P_r(t, \theta) \, =  \,p_r(t) \, + \, \theta  \,\bar{K}(t),
\end{eqnarray}
where the superspace coordinates $(t, \theta)$ characterize the (1, 1)-dimensional
chiral super-submanifold, the supervariables $(\Lambda, F,\bar{F}, P_r)$ are defined on {\it this}
super-submanifold and $(\bar{R}, \bar{B}_1, \bar{B}_2, \bar{K})$ are the secondary
variables. It is self-evident that, in the limit $\theta = 0$, we get back our ordinary
variables $(\lambda, C, \bar{C}, p_r)$ and the pairs $(\bar{B}_1, \bar{B}_2)$ and
$(\bar{R}, \bar{K})$ are bosonic and fermionic in nature, respectively, on the
r.h.s. of expansions (15). This inference is drawn due to the fermionic (${\theta}^2 = 0$)
nature of the Grassmannian variable $\theta$.

We shall proceed in the same manner as has been done in the methodology of  derivation of the BRST symmetry
transformations. Thus, first of all, we note that the following quantities are anti-BRST
invariant, namely;
\begin{eqnarray}
&& s_{ab}\,[r, \vartheta, \, p_\vartheta, \, B, \, \bar{C}] = 0,\quad\qquad s_{ab}\,[\lambda \, \dot{\bar{C}}] = 0,
\quad\qquad\quad s_{ab}\,[p_r  \,\bar{C}] =0,  \\ \nonumber
&& s_{ab}\,[\dot{p_r} \, + \lambda] = 0, \qquad s_{ab}\,[\dot{B} \,\lambda + i \,\dot{\bar{C}} \,\dot{C}] = 0, 
\qquad s_{ab}\,[B\, p_r  \,-  \,i \,\bar{C}\,C] = 0.
\end{eqnarray}
According to the basic tenets of the augmented supervariable approach (AVSA/ACSA), we have the following
non-trivial equalities, namely;
\begin{eqnarray}
&& \bar{F}^{(AB)}(t, \theta) = \bar{C}(t), \qquad \dot P_r(t, \theta) 
+ \Lambda(t, \theta) = \dot{p}_r(t) + \lambda(t), \nonumber\\ 
&& P_r(t, \theta) \,\bar{F}^{(AB)}(t, \theta) \,=\, p_r(t) \,\bar{C}(t), \quad 
\Lambda(t, \, \theta) \,\dot{\bar{F}}^{(AB)}(t, \theta) = \lambda(t) \,\dot{\bar{C}}(t),   \nonumber\\
&& \dot{B}(t, \theta) \,\Lambda(t, \theta)  \,+  \,i \,\dot{\bar{F}}^{(AB)}  \,\dot{F}(t, \theta) =  
\dot{B}(t) \,\lambda(t)  \,+ \, i\,\dot{\bar{C}} \,\dot{C}(t).
\end{eqnarray}
In other words, we demand that the anti-BRST  invariant quantities (16) must be independent of the ``soul" 
coordinate $\theta$ because the latter is {\it not} a ``physical" object [25].

The substitution of the expansions (15) leads to the following very useful relationships
\begin{eqnarray}
&& \bar R(t)\, \dot{\bar C}(t) = 0, \qquad\qquad \bar K(t)\,\bar C(t) = 0,
 \qquad\qquad  \dot {\bar K}(t)  + \bar R(t) = 0, \nonumber\\
&& \dot B(t)\, \bar R(t) - \dot{\bar C}(t)\,\dot{\bar B}_1 (t)\,  = 0, \qquad\qquad  \bar K(t)\, B(t)\, 
- \bar C(t)\,\bar {B_1}(t) \,  = 0,
\end{eqnarray}
where we have taken ${\tilde B}^{(AB)}(t, \theta) = B(t)$. It is straightforward to note that the 
following solutions (modulo a constant overall factor) 
\begin{eqnarray}
\bar R(t) = \dot{\bar C}(t),\quad\qquad \bar K(t) = - \,\bar C(t), \quad\qquad {\bar B}_1 (t) 
= -\, B(t) ,\quad\qquad {\bar B}_2(t) = 0,
\end{eqnarray}
satisfy all the relationships of (18). Thus, we have the following expansions 
\begin{eqnarray}
&&\Lambda^{(AB)}(t, {\theta})\, =\, \lambda(t) \,+ \,{\theta}\,(\dot{\bar C}(t)) \,\equiv  \,\lambda (t)\,+ 
\,{\theta}\, (s_{ab} \, \lambda(t)), \nonumber\\
&&F^{(AB)}(t, \theta)\, =\, C(t) \,+ \,{\theta}\,(-\,i\,B(t)) \,\equiv \, C(t) \, + \, {\theta}\, (s_{ab}\,C(t)),\nonumber\\
&&{\bar{F}}^{(AB)}(t,  {\theta})\, = \,\bar{C}(t)\, +\, {\theta}\, (0)\, \equiv\,  \bar{C}(t) \, 
+\, {\theta}\, (s_{ab}\,\bar{C}(t)),\nonumber\\
&&{P_r}^{(AB)}(t, \theta)\, =\,  p_r (t)\, + \,{\theta}\, (-\, {\bar C}(t))\, \equiv \, p_r (t) \, + {\theta}\, (s_{ab}\,p_r (t)).
\end{eqnarray}
where the superscript $(AB)$ on the supervariables denotes the expansions of supervariables
after the application of the anti-BRST invariant restrictions (17).

A close look at equation (20) demonstrates that we have already obtained the {\it non-trivial}
anti-BRST symmetry transformations (2). The trivial anti-BRST transformations 
$s_{ab}[r, \vartheta, p_{\vartheta}, B] = 0$ are self-evident because
 they are 
same as (8) except the fact that the superscript $(\cal B)$ has to be replaced by $(AB)$ and $\bar\theta \to \theta$. 
We obtain the analogue of (14) as a relationship between 
$\partial_\theta $ and $s_{ab}$ as: $\partial_\theta\,\Omega^{(AB)}(t, {\theta}) = 
\, s_{ab}\,\omega(t)\, \equiv \, \pm\, i\, \bigl[\omega(t), \, Q_{ab} \bigr]_{\pm}$ 
which provides the geometrical interpretation for the nilpotent symmetry transformation $s_{ab}$
on the ordinary generic variable $\omega(t)$ in terms of the translation of the corresponding
supervariable $\Omega^{(AB)}(t, \theta)$ along the Grassmannian direction ($\theta$) of the chiral
super-submanifold. We also observe that nilpotency ($\partial_{\theta}^{2} = 0$) of the translation 
generator $\partial_{\theta}$ implies the nilpotency of the transformation $s_{ab}$ and its
generator $Q_{ab}$. Thus, the operators ($\partial_{\theta},\, s_{ab},\, Q_{ab}$) are
beautifully inter-related with one-another as far as the nilpotency property is concerned.

\section{Nilpotent (Anti-)co-BRST Symmetries: Supervariable  Formalism}

To derive the co-BRST symmetry of (4), we follow the same mathematical procedure
as we have adopted for the derivation of the BRST symmetries in our previous section.
In this connection, we observe that the following quantities are co-BRST invariant, namely;
\begin{eqnarray}
&& s_{d}\,[r,\, \vartheta,\, p_{\vartheta},\, B,\, \bar{C}]\,=\,0,\quad 
s_{d}\,[\lambda\,\bar{C}]\,=\,0,\quad s_{d}\,[p_{r}\,\dot{\bar{C}}]\,=\,0, \nonumber \\
&& s_{d}\,[\dot{\lambda}\,-\,p_{r}]\,=\,0, \quad s_{d}\,[\dot{r}\,p_{r}\, -\,i\,\dot{\bar{C}}\,\dot{C}]\,=
\,0, \quad s_{d}\,[\lambda\,(r-a)\,-\,i\,\bar{C}\,C]\,=\,0.
\end{eqnarray}
As a consequence of the above, we shall obtain the useful co-BRST invariant restrictions 
on the supervariables when we shall generalize the ordinary variables to their 
counterpart supervariables on the (1, 1)-dimensional {\it anti-chiral} supermanifold as is given in (7).

Against the backdrop of the above arguments, we have the following non-trivial
restrictions, according to the basic tenets of the augmented supervariable 
approach
\begin{eqnarray}
&& \bar{F}^{(D)}(t,\, \bar{\theta})\,=\,\bar{C}(t), \qquad\qquad\qquad \Lambda(t,\,\bar\theta)\bar{F}^{(D)}
\,(t,\,\bar{\theta})\,=\, \lambda(t)\,\bar{C}, \nonumber \\
&& P_{r}(t,\,\bar{\theta})\,\dot{\bar{F}}^{(D)}(t,\,\bar{\theta})\, =\,p_{r}(t)\,
\dot{\bar{C}}(t), \qquad
\dot{\Lambda}(t,\,\bar{\theta})\,-\,P_{r}(t,\,\bar{\theta})\,=\,\dot{\lambda}(t)\,-\,p_{r}(t),
\nonumber \\
&& \Lambda(t,\,\bar{\theta})\,(r-a)\,-\,i\,\bar{F}^{(D)}(t,\,\bar{\theta})\,
F(t,\,\bar{\theta})\,=\,\lambda (t)\,(r-a)\,-\,i\,\bar{C}(t)\,C(t), \nonumber \\
&& \dot{r} (t)\,P_{r}(t,\,\bar{\theta})\,-\,i\,\dot{\bar{F}}^{(D)}(t,\,\bar{\theta})\,\dot{F}(t,\,\bar{\theta})
\,=\,\dot{r} (t) \,p_{r}(t)\,-\,i\,\dot{ \bar C}(t)\,\dot C(t), 
\end{eqnarray}  
where we have taken into account $\tilde{R}^{(D)} (t,\,\bar{\theta}) = r(t)$.
The dual-BRST invariant variables $s_{d}\,(r,\,\vartheta,\,p_\vartheta,\,B)\,=\,0$ 
automatically imply similar kind of relations as  given in our earlier equation (8) 
except the fact that we have to replace the superscript $(\cal B)$ by $(D)$.
The above restrictions lead to the following relationships between the secondary variables
and the dynamical (as well as auxiliary variables) of our present theory [cf. (1)], namely;
\begin{eqnarray}
&& R(t)\,\bar{C}(t)\,=\,0,\quad\qquad\,\, K(t)\,\dot{\bar{C}}(t)\,=\,0,\quad\qquad\,\, \dot{R}(t)\,=\,K(t),\nonumber \\
&& \dot{\bar{C}}(t)\, \dot{B}_1(t)\, - \,K(t)\,\frac{d}{dt}(r-a)\,=\,0, \quad\qquad B_{1} (t) \bar{C} (t) \,-\,
(r-a)\,R(t)\,=\,0.
\end{eqnarray}
It is straightforward to check that the following solutions (modulo a constant overall factor)
\begin{eqnarray}
R(t)\,=\,\bar{C}(t),\quad\qquad K(t)\,=\,\dot{\bar{C}}(t),\quad\qquad B_{1}(t)\,=\,(r-a),
\end{eqnarray}
satisfy {\it all} the relationships that are given in (23).
The substitution of (24) into the expansion (7) leads to the following:  
\begin{eqnarray}
&& \Lambda^{(D)}(t,\,\bar{\theta})\,=\,\lambda(t)\,+\,\bar{\theta}\,(\bar{C}(t))\,
\equiv\,\lambda(t)\,+\,\bar{\theta}\,(s_{d}\,\lambda(t)), \nonumber \\
&& F^{(D)}(t,\,\bar{\theta})\,=\,C(t)\,+\,\bar{\theta}\,(i\,[r-a])\,\equiv \,
C(t)\,+\,\bar{\theta}\,(s_{d}\,C(t)), \nonumber \\
&& \bar{F}^{(D)}(t,\,\bar{\theta})\,=\,\bar{C}(t)\,+\,\bar{\theta}\,(0)\,\equiv\,
\bar{C}(t)\,+\,\bar{\theta}\,(s_{d}\,\bar{C}(t)), \nonumber \\
&& P_{r}^{(D)}(t,\,\bar{\theta})\,=\,p_{r}(t)\,+\,\bar{\theta}\,(\dot{\bar{C}} (t))\,
\equiv \, p_{r}(t)\,+\,\bar{\theta}\,(s_{d}\,p_{r} (t)).
\end{eqnarray}   
A close  look at (25) and the analogue of (8) demonstrates that we have already derived 
the co-BRST symmetry transformation (4). The above expansion in (25) leads to the
geometrical interpretation of $s_{d}$, too. In other words,
from the above equation, it is clear that we have a relationship which is exactly like (14) 
except the fact the superscript on the supervariables would be $(D)$ and the replacements: $s_b \to s_d, Q_b \to Q_d$. 
Thus, dual-BRST (or co-BRST) symmetry $s_d$ acting on a generic ordinary variable is 
equivalent to the translation of the corresponding supervariable (that is obtained 
after the application of the co-BRST invariant restriction) along the
$\bar\theta$-direction of the anti-chiral super-submanifold.

We concentrate now on the derivation of the anti-co-BRST symmetries, for which, the expansions (15) for 
the supervariables (on the (1, 1)-dimensional {\it chiral} super-submanifold) are to be taken into account.
The most decisive feature of our augmented supervariable approach is to find out the invariant quantities.
 It is gratifying to state that we have found out the following useful anti-co-BRST invariant quantities:
\begin{eqnarray}
&& s_{ad}\,[r,\, \vartheta,\, p_{\vartheta},\, B,\,{C}]\,=\,0,\quad 
s_{ad}\,[\lambda\,{C}]\,=\,0,\quad s_{ad}\,[p_{r}\,\dot{C}]\,=\,0, \nonumber \\
&& s_{ad}\,[\dot{\lambda}\,-\,p_{r}]\,=\,0, \quad s_{ad}\,[\dot{r}\,p_{r}\, -\,i\,\dot{\bar{C}}\,\dot{C}]\,=
\,0, \quad s_{ad}\,[\lambda\,(r-a)\,-\,i\,\bar{C}\,C]\,=\,0.
\end{eqnarray}
According to the basic rules of the augmented supervariable formalism, we have
the following anti-co-BRST restrictions on the supervariables:
\begin{eqnarray}
&& {F}^{(AD)}(t,\, {\theta})\,=\,{C}(t), \quad\qquad\quad\Lambda(t,\,\theta)\,{F}^{(AD)}
\,(t,\,{\theta})\,=\, \lambda(t)\,{C}(t), \nonumber \\
&& P_{r}(t,\,{\theta})\,{F}^{(AD)}(t,\,{\theta})\, =\,p_{r}(t)\,{C}(t), \quad\quad
\dot{\Lambda}(t,\,{\theta})\,-\,P_{r}(t,\,{\theta})\,=\,\dot{\lambda}(t)\,-\,p_{r}(t),
\nonumber\\
&& \Lambda(t,\,{\theta})\,(r-a)\,-\,i\,\bar{F}(t,\,{\theta})\,
F^{(AD)}(t,\,{\theta})\,=\,\lambda(t)\,(r-a)\,-\,i\,\bar{C}(t)\,C(t), \nonumber \\
&& \dot{r} (t)\,P_{r}(t,\,{\theta})\,-\,i\,\dot{\bar{F}}(t,\,{\theta})\,{\dot{F}}^{(AD)}(t,\,{\theta})
\,=\,\dot{r} (t) \,p_{r}(t)\,-\,i\,\dot{\bar C}(t)\,\dot C(t).
\end{eqnarray} 
where we have taken $r(t) \rightarrow \tilde{R}^{(AD)}(t, \bar\theta) = r(t)$ and the superscript $(AD)$ on $F(t, \theta)$
demonstrates that this supervariable has {\it no} expansion along $\theta$-direction of the chiral super-submanifold.
This is due to the fact that the ghost variable $C(t)$ is an anti-dual-BRST (i.e. anti-co-BRST) invariant quantity 
(i.e. $s_{ad}\, C = 0$). The above relationships  (27) lead to the following very useful relations
\begin{eqnarray}
&& \bar R(t)\,{C}(t)\,=\,0,\quad\qquad \bar K(t)\,\dot{{C}}(t)\,=\,0,\quad\qquad \dot{\bar R}(t)\,=\,\bar K(t),\nonumber \\
&& \bar{B}_2(t)\,C(t)\,+\,(r-a)\,\bar R(t) = 0, \quad\qquad \dot{\bar B}_{2}(t)\,\dot{C}(t)\,
+\,\frac{d}{dt}\,(r-a)\,\bar K(t) = 0,
\end{eqnarray}  
which allow us to choose  the following solutions (modulo an overall constant  factor):
\begin{eqnarray}
\bar R(t)\,=\,{C}(t),\;\quad  \bar K(t)\,=\,\dot{C}(t),\;\quad {\bar B}_{2}(t)\,=\,-\,
\bigl (r (t) - a \bigr ), \;\quad {\bar B}_1(t)\,=\, 0.
\end{eqnarray}
The substitution of the above expressions for the secondary variables into the super-expansion (15)
leads to the following
\begin{eqnarray}
&& \Lambda^{(AD)}(t,\,\theta)\,=\,\lambda(t)\,+\,\theta\,(C(t))\,\equiv \,
\lambda(t)\,+\,\theta\,(s_{ad}\,\lambda(t)), \nonumber \\
&& F^{(AD)}(t,\,\theta)\,=\,C(t)\,+\,\theta(0)\equiv \, C(t)\,+\,\theta\,(s_{ad}\,C(t)), \nonumber \\
&& \bar{F}^{(AD)}(t,\,\theta)\,=\,\bar{C}\,+\,\theta (-\,i\,(r-a))\equiv \bar{C}(t)\,
+\,\theta\,(s_{ad}\,\bar{C}(t)), \nonumber \\
&& P_{r}^{(AD)}(t,\,\theta)\,=\,p_{r}(t)\,+\,\theta\,(\dot{C})\equiv p_{r}(t)\,
+\,\theta\,(s_{ad}\,p_{r}(t)),
\end{eqnarray}
where the superscript $(AD)$ denotes the expansions of the supervariables after the application 
of the anti-co-BRST invariant restrictions quoted in (27). The trivial anti-co-BRST invariance of some variables 
[e.g. $s_{ad} (r,\,\vartheta,\,p_{\vartheta},\,B) = 0]$ implies that we have the analogue of 
equation (8) with the replacement $({\cal B}) \rightarrow (AD)$ and $\bar\theta \rightarrow \theta$.
 A close and careful
observation of the above expansions show that we have already obtained the
non-trivial anti-co-BRST symmetry transformations $(s_{ad})$ of (4). The above relation (30) also 
provides the geometrical interpretation of $s_{ad}$ in the language  of the translational
generator $\partial_\theta$ on the  (1, 1)-dimensional {\it chiral} super-submanifold of the general
(1, 2)-dimensional supermanifold (on which our 1D theory of Hodge theory (i.e. rigid rotor) is considered).
To be more explicit, we have the mapping $s_{ad} \leftrightarrow \partial_\theta $. In other words, the translation of the
chiral {\it supervariables} (obtained after the anti-co-BRST invariant restrictions (27)) along
the Grassmannian direction $\theta$ of the {\it chiral} super-submanifold produces the anti-co-BRST symmetry
transformation ($s_{ad}$) for the corresponding {\it ordinary} variable.

\section{Invariance of Lagrangian: Supervariable  Technique}

In this section, we concentrate on the symmetry properties of the Lagrangian (cf. Sec. 2)
within the framework of the augmented (anti-)chiral supervariable approach to BRST
formalism. Towards this goal in mind, 
we can capture the (anti-)BRST and (anti-)co-BRST invariance of the starting Lagrangian (1)
in the language of the augmented (anti-)chiral supervariable approach. To this end in mind, we observe
that the Lagrangian (1) can be generalized onto the (1, 1)-dimensional (anti-)chiral super-submanifolds
(of the {\it general} (1, 2)-dimensional supermanifold) as:
\begin{eqnarray}
L_{b}\rightarrow \tilde{L}_{b}^{(\cal B)}\,&=&\,\dot{r}\,P_{r}^{(\cal B)}(t,\,\bar{\theta})\,
+\,\dot\vartheta\,p_{\vartheta}\,-\,\frac{p_{\vartheta}^{2}}{2 r^2}\,-\,
\Lambda^{(\cal B)}t,\,\bar{\theta})\,(r-a)\,+\,\frac{B^{2}(t)}{2} \nonumber \\
& \, +\,& B(t)\,\bigl[\dot{\Lambda}^{(\cal B)}(t,\bar{\theta})\,-\,P_{r}^{(\cal B)}(t,\,\bar{\theta})\bigr]
\,-\,i\,\dot{\bar{F}}^{(\cal B)}(t,\,\bar{\theta})\,\dot{C}(t) \nonumber \\
&\,+\,& i\,\bar{F}^{(\cal B)}(t,\,\bar{\theta})\,C(t),\nonumber\\
L_{b}\rightarrow \tilde{L}_{b}^{(D)}\,& = &\,\dot{r}\,P_{r}^{(D)}(t,\,\bar{\theta})\,
+\,\dot\vartheta\,p_{\vartheta}\,-\,\frac{p_{\vartheta}^{2}}{2 r^2}\,-\,
\Lambda^{(D)}(t,\,\bar{\theta})\,(r-a)\,+\,\frac{B^{2}(t)}{2} \nonumber \\
& \, +\,& B(t)\,\bigl[\dot{\Lambda}^{(D)}(t,\bar{\theta})
\,-\,P_{r}^{(D)}(t,\,\bar{\theta})\bigr]\,-\,i\,\dot{\bar{C}}(t)\,\dot{F}^{(D)}(t,\,\bar{\theta})
\nonumber \\
&\,+\,& i\,\bar{C}(t)F^{(D)}(t,\,\bar{\theta}),\nonumber\\
L_{b}\rightarrow \tilde{L}_{b}^{(AB)}\,& = &\,\dot{r}\,P_{r}^{(AB)}(t,\,{\theta})\,
+\,\dot\vartheta\,p_{\vartheta}\,-\,\frac{p_{\vartheta}^{2}}{2 r^2}\,+\,
\Lambda^{(AB)}(t,\,\bar{\theta})\,(r-a)\,+\,\frac{B^{2}(t)}{2} \nonumber \\
& \, +\,& B(t)\,\bigl[\dot{\Lambda}^{(AB)}(t,{\theta})\,-\,P_{r}^{(AB)}(t,\,{\theta})\bigr]
\,-\,i\,\dot{\bar{C}}\,\dot{F}^{(AB)}(t,\,\theta) \nonumber \\
&\,+\,& i\,\bar{C}(t){F}^{(AB)}(t,\,{\theta}),\nonumber\\
L_{b}\rightarrow \tilde{L}_{b}^{(AD)}\,& = &\,\dot{r}\,P_{r}^{(AD)}(t,\,{\theta})\,
+\,\dot\vartheta\,p_{\vartheta}\,-\,\frac{p_{\vartheta}^{2}}{2 r^2}\,+\,
\Lambda^{(AD)}(t,\,{\theta})\,(r-a)\,+\,\frac{B^{2}(t)}{2} \nonumber \\
& \, +\,& B(t)\,\bigl[\dot{\Lambda}^{(AD)}(t,{\theta})\,-\,P_{r}^{(AD)}(t,\,{\theta})\bigr]
\,-\,i\,\dot{\bar{F}}^{(AD)}(t,\,{\theta})\,\dot{C}(t) \nonumber \\
&\,+\,&i\,\bar{F}^{(AD)}(t,\,{\theta})\,C(t),
\end{eqnarray}
where the super-Lagrangians have been denoted by $ \tilde{L}_{b}^{(r)} (r = {\cal B}, AB, D, AD)$ and the
superscripts on the supervariables denote the nature of the supervariables 
(that have been obtained after the application of (anti-)BRST and (anti-)co-BRST
restrictions). These restrictions have been made transparent and lucid in the main body
of our text. The above forms of the Lagrangians incorporate some supervariables {\it plus} some 
terms that are ordinary variables (because the latter are invariant quantities). For instance,
the variables ($ r, \vartheta, p_{\vartheta}, B$) have been taken to be {\it ordinary}
everywhere because they are (anti-)BRST as well as (anti-) co-BRST invariant 
(i.e. $s_{(a)b} [r, \vartheta, p_{\vartheta}, B] = 0$ and $s_{(a)d} [r, \vartheta, p_{\vartheta}, B] = 0$).

It is elementary to check that the following are true, namely; 
\begin{eqnarray}
&&\frac{\partial}{\partial\,{\bar\theta}}\, \tilde{L}_{b}^{(\cal B)}\,= 
\frac{d}{dt}\, \bigl[ B\,\dot C - (r-a)\, C \bigr] \equiv \, s_b \, L_b,\nonumber\\
&&\frac{\partial}{\partial\,{\theta}}\, \tilde{L}_{b}^{(AB)}\,= \,
\frac{d}{dt}\, \bigl[ B\,\dot {\bar C} - (r-a)\, \bar C \bigr] \equiv \, s_{ab} \, L_b,\nonumber\\
&&\frac{\partial}{\partial\,{\bar\theta}}\, \tilde{L}_{b}^{(D)}\,= \,
\frac{\partial}{\partial\,{\theta}}\, \tilde{L}_{b}^{(AD)}=\, 0\, \equiv \, s_{(a)d}\, L_b,
\end{eqnarray}
due to the mappings, shown in the main body of our text which demonstrate that:
$\partial_{\bar{\theta}}\Longleftrightarrow (s_{b},\,s_{d})$ and
$\partial_{\theta}\Longleftrightarrow (s_{ad},\,s_{ab})$. The above relations also 
provide the geometrical basis for the (anti-)BRST and (anti-)co-BRST invariance of the 
starting Lagrangian (1). It states that the super Lagrangians 
$\tilde{L}_{b}^{(r)} (r\,=\,{\cal B}, AB, D, AD)$ are expressed in terms of the composite
(super)variables that are $\it{either}$ present in the starting Lagrangian (1)
and/or obtained after the application of the (anti-)BRST and (anti-)co-BRST
invariant restrictions. These sum of terms in the super Lagrangian 
$\tilde{L}_{b}^{(r)} (r\,=\,{\cal B}, AB, D, AD)$ are such that their translations
along the Grassmannian $(\theta,\,\bar{\theta})$-directions of the (anti-)chiral 
super sub-manifolds $\it{either}$ lead to zero result (i.e. (anti-)co-BRST invariance)
$\it{or}$ total time derivatives (i.e. (anti-) BRST invariance). 
In either cases, the action integral remains invariant.

\section{Nilpotency  and Absolute Anticommutativity: \\Supervariable Method}

In this section, we dwell on the two key and clinching properties of BRST symmetries 
in the terminology of augmented (anti-)chiral supervariable approach. In this 
context, first of all, we now focus on the nilpotency property of the conserved 
(anti-)BRST and (anti-)co-BRST charges within in the framework of augmented 
ACSA to BRST formalism. It can be readily shown that these charges can be 
expressed in terms of the Grassmannian differentials, derivatives, integrals  
and (super)variables as follows
\begin{eqnarray*}
Q^{(1)}_{d} &\,=&\,-\, i\,\frac{\partial}{\partial\bar{\theta}}\big[\dot{{F}}^{(D)}(t,\,\bar{\theta})\,
\bar{F}^{(D)}(t,\,\bar{\theta}) +\,i\,\dot{\Lambda}^{(D)}(t,\,{\bar\theta})\,\dot{B}(t)\big]\,\nonumber\\
&\equiv & -\,i\, \int \, d\bar\theta \,\big[\dot{{F}}^{(D)}(t,\,\bar{\theta})\,
\bar{F}^{(D)}(t,\,\bar{\theta}) +\,i\,\dot{\Lambda}^{(D)}(t,\,{\bar\theta})\,\dot{B}(t)\big],\,\nonumber\\
Q^{(2)}_d &=& i\, \frac{\partial}{\partial{\theta}} \big[\dot{{\bar F}}^{(AD)}(t,\,{\theta})\,
{\bar F}^{(AD)}(t,\,{\theta})\bigr]\quad
\equiv  \,i\, \int \, d\theta \,\big[\dot{{\bar F}}^{(AD)}(t,\,{\theta})\,{\bar F}^{(AD)}(t,\,{\theta})\bigr],\nonumber\\
Q^{(1)}_{ad}&\,=\,& i\,\frac{\partial}{\partial{\theta}}\big[\,\dot{\bar{F}}^{(AD)}(t,\,{\theta})\, {F^{(AD)}} \,(t,\,{\theta})\,
-\,i\,\dot{\Lambda}^{(AD)}(t,\,{\theta})\,\dot{B}(t)\big] \nonumber \\
&\equiv & \,i\, \int \, d\theta \,\big[\,\dot{\bar{F}}^{(AD)}(t,\,{\theta})\, {F^{(AD)}} \,(t,\,{\theta})\,
-\,i\,\dot{\Lambda}^{(AD)}(t,\,{\theta})\,\dot{B}(t)\big], \nonumber \\
Q^{(2)}_{ad}&=&\,-\,i\, \frac{\partial}{\partial\bar{\theta}} \big[\dot{{F}}^{(D)}(t,\,\bar{\theta})\,
{F}^{(D)}(t,\,{\bar\theta})\bigr]\quad
\equiv  -\,i\, \int \, d\bar\theta \,\big[\dot{{F}}^{(D)}(t,\,\bar{\theta})\,{F}^{(D)}(t,\,{\bar\theta})\bigr],\nonumber\\
\end{eqnarray*}
\begin{eqnarray}
Q^{(1)}_{b} &\,=\,&-\, i\,\frac{\partial}{\partial\bar{\theta}}\big[{F}^{({\cal B})}(t,\,\bar{\theta})\,
\dot{\bar F}^{({\cal B})}(t,\,\bar{\theta}) +\, i\,B(t)\,\Lambda^{({\cal B})} (t, \bar\theta)\big]\nonumber\\
&\equiv & -\,i\, \int \, d\bar\theta \,\big[{F}^{({\cal B})}(t,\,\bar{\theta})\,
\dot{\bar F}^{({\cal B})}(t,\,\bar{\theta}) +\, i\,B(t)\,\Lambda^{({\cal B})} (t, \bar\theta)\big],\nonumber\\
Q^{(2)}_b\,&=&-\,i\,\frac{\partial}{\partial{\theta}}\big[{\dot F}^{(AB)}(t, \theta)\,{F}^{(AB)}(t,\theta)\bigr]\quad
\equiv -\,i\, \int \, d\theta \,\big[{\dot F}^{(AB)}(t, \theta)\,{F}^{(AB)}(t,\theta)\bigr],\nonumber\\
Q^{(1)}_{ab} &\,=\,&\frac{\partial}{\partial{\theta}}\big[B\,\Lambda^{(AB)}(t, \theta) +\,i\,{\bar F}^{(AB)}(t,\,{\theta})\,
{\dot F}^{(AB)}(t,\,{\theta})\big]\nonumber\\
&\equiv & \, \int \, d\theta \,\big[B\,\Lambda^{(AB)}(t, \theta) +\,i\,{\bar F}^{(AB)}(t,\,{\theta})\,
{\dot F}^{(AB)}(t,\,{\theta})\big],\nonumber\\
Q^{(2)}_{ab}&=&i\, \frac{\partial}{\partial{\bar\theta}}\big[\dot{\bar F}^{({\cal B})}(t, \bar\theta)\,
{\bar F}^{({\cal B})}(t,\bar\theta)\bigr]\quad
\equiv  i\, \int \, d\bar\theta \,\big[\dot{\bar F}^{({\cal B})}(t, \bar\theta)\,{\bar F}^{({\cal B})}(t,\bar\theta)\bigr],
\end{eqnarray}  
where, in the above, we have used the equations of motion: $\ddot B + B = 0,\, \dot B + (r - a) = 0$
in the derivation of {\it two} types of expressions for the (anti-)co-BRST charges $Q_{(a)d}$. It is clear, 
from the above expressions, that $\partial_\theta Q^{(1)}_r = 0 \,\, (r = ab, ad)$ and 
$\partial_{\bar \theta} Q^{(1)}_r = 0\,\, (r = b, d)$ due to the nilpotency ($\partial_\theta^2 =
 \partial_{\bar\theta}^2 =0 $) of the translational generators ($\partial_\theta, \partial_{\bar\theta}$) 
along the Grassmannian directions ($\theta, \bar\theta$). In other words, the nilpotency of the (anti-)BRST
 and (anti-)co-BRST charges are deeply connected with the nilpotency  ($\partial_\theta^2 = 
\partial_{\bar\theta}^2 =0 $) of the translational generators ($\partial_\theta, \partial_{\bar\theta}$)
along the Grassmannian directions.

The above statements become more transparent when we express the conserved charges in terms of the nilpotent
transformations and ordinary fields in the ordinary space. For instance, we have the following expressions for
the charges in the ordinary space, namely; 
\begin{eqnarray}
&& Q_d = -\, i\, s_d \,[i\, {\dot B}\, {\dot \lambda} + \dot{ C} \, {\bar C}] = \,i\, s_{ad}\, 
[\dot{\bar C}\, {\bar C}],\nonumber\\ 
 &&Q_{ad} =\, i\, s_{ad}\, [-\,i\, {\dot B}\, {\dot \lambda} + \dot{ \bar C} \, {C}] = -\,\,i\, s_{d}\,[\dot{C}\, { C}],\nonumber\\
&&Q_b = -\, i\, s_b \,[i\, B\, \lambda - \dot{\bar C} \, C] = \,-\,i\, s_{ab}\, [{\dot C}\, C],\nonumber\\
&& Q_{ab} =\,  s_{ab} \,[B\, \lambda + i\, {\bar C} \, {\dot C}] = \,i\,s_{b}\, [\dot {\bar C}\, {\bar C}],
\end{eqnarray} 
 due to the mappings: $\partial_\theta \leftrightarrow (s_{ab}, s_{ad})$ and  
$\partial_{\bar \theta} \leftrightarrow (s_{b}, s_{d})$
 and the limiting cases of the superfields (when $\theta = \bar\theta = 0$). It is very clear that
\begin{eqnarray}
&& s_{d}\,Q_{d}\, = \,i\,\lbrace{Q_{d},\,Q_{d}}\rbrace\,=\,0, \quad \quad \quad\,\,\, \,
s_{ad}\,Q_{d}\, = \,i\,\lbrace{Q_{d},\,Q_{ad}}\rbrace\, = \,0,\nonumber\\
&& s_{ad}\,Q_{ad}\, = \,i\,\lbrace{Q_{ad},\,Q_{ad}}\rbrace\, = \,0, \quad \quad \,
s_{d}\,Q_{ad}\,= \,i\,\lbrace{Q_{ad},\,Q_{d}}\rbrace\, = \,0, \nonumber\\ 
&& s_{b}\,Q_{b}\, = \,i\,\lbrace{Q_{b},\,Q_{b}}\rbrace\, = \,0, \qquad \qquad 
s_{ab}\,Q_{b}\, = \,i\,\lbrace{Q_{b},\,Q_{ab}}\rbrace\, = \,0, \nonumber\\
&& s_{ab}\,Q_{ab}\, = \,i\,\lbrace{Q_{ab},\,Q_{ab}}\rbrace\, = \,0, \quad \quad\,\,\,
s_{b}\,Q_{ab}\, = \,i\,\lbrace{Q_{ab},\,Q_{b}}\rbrace\, = \,0, 
\end{eqnarray}
where we have exploited the standard definition of a {\it generator} for a given {\it continuous} symmetry 
transformation. The crucial point is to note that the properties of the nilpotency and  
absolute anticommutativity are inter-connected in a beautiful fashion. We stress that the 
above relations have been obtained due to our knowledge of the supervariable formalism.

We end this section with the remark that it is a {\it novel} observation for us that absolute anticommutativity
of the (anti-)BRST and (anti-)co-BRST charges emerge {\it even} when we take the (anti-)chiral expansions for
the (anti-)chiral supervariables (that are defined on the (1, 1)-dimensional super-submanifolds of the general
(1, 2)-dimensional supermanifold on which our ordinary theory 
of a 1D rigid rotor is generalized). This happens, we speculate, because of the
fact that the nilpotency ($\partial_\theta^2 = \partial_{\bar\theta}^2 = 0 $) property is {\it basically}
a limiting case of the absolute anticommutativity ($\partial_\theta \partial_{\bar\theta} 
+ \partial_{\bar\theta} \partial_{\theta} =0 $)
property of the translational generators ($\partial_\theta, \partial_{\bar\theta} $) when we take the limit
$\partial_\theta = \partial_{\bar\theta}$ (and/or $\partial_{\bar\theta} = \partial_\theta$).

\section{Conclusions}

One of the key results of our present investigation is the observation that one 
can avoid the {\it mathematical} use of (dual-)horizontality conditions and replace 
it by the physically motivated (anti-)BRST as well as (anti-)co-BRST 
invariant restrictions (within the framework of the augmented version of {\it (anti-)chiral} 
supervariable approach to BRST formalism) to derive the nilpotent (anti-)BRST and (anti-)co-BRST symmetry 
transformations. Our method of derivation is very simple and intuitive in the sense that 
it provides the geometrical interpretations for the symmetry invariance and nilpotency in
a straightforward manner (which are consistent with the results of theoretical methods 
where the {\it mathematical} strength of (D)HCs are exploited extensively). To be precise, our results lend 
support to the results obtained by the precise mathematical application of the (D)HCs [13,27].

The derivation of the (anti-)co-BRST symmetries is a novel result in our present investigation
 because we have {\it not} used the (anti-)co-BRST invariance in our earlier work [27] on the 
supervariable approach to a rigid rotor. Rather, we have exploited 
the virtues of the dual-horizontality condition to derive {\it all} the above symmetry transformations.
We are unable to derive the analogue of the Curci-Ferrari condition [26] within the framework of 
our present approach. However, the latter condition is a {\it trivial} relationship in our present 
theory because we obtain the absolutely anticommuting and nilpotent (anti-)BRST and (anti-)co-BRST 
symmetry transformations in a straightforward fashion for a {\it single} Lagrangian.

In our very recent works [28-30], we have exploited the idea of ACSA to BRST formalism in the context of (non-)Abelain 1-form
gauge theories where there is interaction between matter and gauge fields. We have been able to derive
{\it all} the (anti-)BRST symmetry transformations by exploiting the (anti-)BRST invariant
restrictions on the (anti-)chiral superfields. Further, we have been able to establish the 
nilpotency and absolute anticommutativity of the
conserved fermionic charges where the Curci-Ferrari condition [26] has played a pivotal role. 
To be precise, in the proof of the absolute anticommutativity of the (anti-)BRST charges, we have taken the
help of the Curci-Ferrari restriction [26] for the interacting non-Abelian 1-form gauge theory with Dirac fields.
We have also considered the 2D (non-)Abelian gauge theories {\it without} any interaction with matter fields
and applied the ACSA to BRST formalism to derive the (anti-)BRST and (anti-)co-BRST symmetry transformations by 
exploiting the theoretical strength of the (anti-)BRST and (anti-)co-BRST invariant restrictions on the
(anti-)chiral superfields of the theory. We have {\it also} been able to prove the 
nilpotency and absolute anticommutativity of the conserved (anti-)BRST as well as (anti-)co-BRST charges within the 
framework of ACSA to BRST formalism.

In the standard superfield formalism [1-12], the absolute anticommutativity of the 
nilpotent (anti-)BRST symmetry transformations (corresponding to an ordinary 
D-dimensional theory) is guaranteed due to the {\it full} expansion of the superfields  
along {\it all} the Grassmannian directions (i.e. $1, \theta, \bar\theta, \theta\bar\theta$) of the 
(D, 2)-dimensional supermanifold (see, Appendix A). However, we note that, in our present investigation, 
the absolute anticommutativity of the (anti-)BRST and (anti-)co-BRST symmetry generators 
turns up very beautifully (cf. Sec. 6) {\it even} though we have considered the (anti-)chiral 
expansions of the  (anti-)chiral supervariables. This is a completely {\it novel} observation (where we obtain 
the absolute anticommutativity of charges  {\it without} the {\it full} expansion of the 
relevant supervariables of our theory). This observation is one of the highlights of our present investigation.

We have exploited the strength of symmetry invariance in the supersymmetric (SUSY) theories as well
[14-17].  In these works [14-17], we have derived the nilpotent ${\mathcal N} = 2$ SUSY transformations by imposing
the SUSY invariant restrictions on the (anti-)chiral supervariables of the (1, 1)-dimensional 
super-submanifolds of the general (1, 2)-dimensional supermanifold. 
It would be a challenging problem for us to apply our present theoretical technique (based on symmetry invariance) to 
other physically interesting systems that are associated with the gauge and SUSY theories.
We are presently very intensively involved with these issues and our results would be reported in our future 
publications [31].\\

\section*{Acknowledgments}
Two of us (TB and NS) would like to gratefully acknowledge the financial support from BHU-fellowship
under which the present work has been carried out.\ TB would like  to thank the HoD, Physics Department BHU, 
for his kind invitation and warm hospitality at the Department where a part of this work was completed.\\

\begin{center}
{\bf Appendix A: On the Absolute Anticommutativity Property}\\
\end{center}

\noindent
The (anti-)BRST and (anti-)co-BRST symmetry transformations (and corresponding generators) 
{\it always} turn out to be absolutely anticommuting so that their linear independence could 
be encoded in the language of the algebraic relationships. In the standard superfield formalism, a
given superfield  is {\it fully} expanded along ($\theta, \bar\theta$)-directions of 
the $(D, 2)$-dimensional supermanifold, as (see, e.g. [1-12])
\begin{eqnarray}
\Phi(x, \theta, \bar\theta) = \phi(x) + \theta \, \Bar  L (x) + \bar\theta \,  L (x) + i\, \theta\, \bar\theta\,  M (x),
\end{eqnarray}
where $\Phi(x, \theta, \bar{\theta})$ is a generic superfield that can be bosonic {\it or}
fermionic in nature and $Z^{M} = (x^{\mu}, \theta, \bar{\theta})$ are the superspace coordinates.
If $\Phi (x, \theta, \bar{\theta})$ were bosonic, it is evident that $ M (x)$  and $\phi(x)$
would be bosonic and $\bigl (L (x),\,\bar {L}(x) \bigr )$ would be fermionic secondary fields. On the 
contrary, if $\Phi(x, \theta, \bar{\theta})$ were fermionic superfield, the pair $\bigl(\phi(x), M (x)\bigr)$
would be fermionic and  the other pair $\bigl(L(x), \bar{L}(x)\bigr)$ would be bosonic secondary fields.
These inferences have been drawn due to the {\it fermionic} nature of the Grassmannian variables 
$(\theta, \bar{\theta})$ of the superspace coordinates which satisfy the standard relationships: ${\theta}^2 = {\bar\theta}^2 = 0, \theta \, \bar\theta
+ \bar\theta \, \theta = 0$.

It is now clear that the following is true, namely:
\begin{eqnarray}
&& \frac{\partial}{\partial{\theta}}\,\frac{\partial}{\partial{\bar{\theta}}}\,\,
\Phi(x, \theta, \bar{\theta}) = - \,i\,M (x), \qquad
\frac{\partial}{\partial{\bar\theta}}\,\frac{\partial}{\partial{\theta}}\,\,
\Phi(x, \theta, {\bar\theta}) = +\, i\,M (x).
\end{eqnarray}
Thus, we have [due to  
$(\partial_{\bar{\theta}}\,\partial_{\theta}\, + \partial_{\theta}\,\partial_{\bar{\theta}})\,\Phi(x, \theta, \bar\theta)\, =\,0$] 
the following mappings (in the operator forms) if we identify
$\partial_{\bar{\theta}}\leftrightarrow(s_{b}, s_{d})$ and $\partial_{\theta}\leftrightarrow
(s_{ab}, s_{ad})$:
\begin{eqnarray}
&& (\partial_{\bar{\theta}}\,\partial_{\theta}\, + \partial_{\theta}\,\partial_{\bar{\theta}}\, =\,0)
\Longleftrightarrow  (s_{b}\, s_{ab} + s_{ab}\, s_{b}\,=\,0,\,\quad s_b\,s_{ad} + s_{ad}\,s_b = 0),  \nonumber \\
&& (\partial_{\bar{\theta}}\,\partial_{\theta}\, + \partial_{\theta}\,\partial_{\bar{\theta}}\, =\,0)
\Longleftrightarrow  (s_{d}\, s_{ad} + s_{ad}\, s_{d}\,=\,0, \quad s_d\,s_{ab} + s_{ab}\,s_d = 0).
\end{eqnarray}
The above mappings demonstrate the absolute anticommutativity of $s_{(a)b}$ and $s_{(a)d}$. Thus,
it is crystal clear that when we take the {\it full} super expansion of the superfield along all the 
$( 1, \theta, \bar\theta, \theta\bar\theta)$-directions of the $(D, 2)$-dimensional supermanifold, we automatically
obtain the absolutely anticommuting (anti-)BRST and (anti-)co-BRST symmetry transformations.     
In the present endeavor, however, we note that we have taken {\it only} the (anti-)chiral super expansions
for the (anti-)chiral supervariables on the (anti-)chiral super-submanifolds of the {\it general} supermanifold
and {\it still} we observe the validity of anticommutativity property.

A close look at the expressions for the conserved and nilpotent charges $Q_{(a)b}$ and  $Q_{(a)d}$ (cf. Sec. 6)
demonstrates that the anticommutativity property of the (anti-)BRST and (anti-)co-BRST charges is hidden 
in these expressions (which are derived due to our knowledge of the augmented supervariable formalism). In fact, it is
worth noting that all the four  fermionic charges have been expressed in terms of the 
derivatives w.r.t. $\theta$ and $\bar\theta$. As a consequence, we note that 
 $\partial_\theta \, Q_r^{(1)} = 0 \,\, (r = ab, ad) $ and  $\partial_{\bar\theta} \, Q_r^{(1)} = 0 \,\, (r = b, d) $.
Furthermore, we also note that $\partial_{\bar\theta} \, Q_r^{(2)} = 0 \,\, (r = ab, ad) $ and
$\partial_{\theta} \, Q_r^{(2)} = 0 \,\, (r = b, d) $. The latter relations capture the anticommutativity property of 
$Q_{(a)b}$ and $Q_{(a)d}$ if we identify $\partial_\theta \leftrightarrow (s_{ab}, \,\, s_{ad})
\leftrightarrow (Q_{ab}, \,\, Q_{ad})$ and $\partial_{\bar\theta} \leftrightarrow (s_{b}, \,\, s_{d}) 
\leftrightarrow (Q_{b}, \,\, Q_{ab})$. The absolutely anticommuting property becomes very transparent and lucid
in the ordinary space where we have the validity of relations (34) and (35). We re-emphasize that it is a {\it novel} observation
that we are able to show the absolutely anticommuting property of the (anti-)BRST and (anti-)co-BRST charges 
{\it even} though we have the (anti-)chiral expansions of the (anti-)chiral supervariables on the 
appropriately chosen (1, 1)-dimensional (anti-)chiral super-submanifolds of the
{\it general} (1, 2)-dimensional supermanifold on which our 1D toy model of the Hodge theory is generalized.\\

\begin{center}
{\bf Appendix B: On ${\mathcal N} = 2$ SUSY Conserved and Nilpotent Charges for a Free Particle
and Comment on their Absolute Anticommutativity Property}\\
\end{center}

\noindent
Before we commence the discussions on the central theme of our present Appendix, we would like to mention that, in general, the
${\mathcal N} = 2$ supersymmetries may {\it not} be nilpotent of order of two. As a consequence, the superfield/supervariable 
approach can {\it not} be applied to such kinds of ${\mathcal N} = 2$ SUSY systems where the SUSY symmetries are not nilpotent
of order two. However, in our present Appendix, we take a simple case of an ${\mathcal N} = 2$ SUSY model of a free particle
where the symmetries are nilpotent of order two. In fact, we have applied ACSA to many SUSY models which are more involved than a 
simple case of an ${\mathcal N} = 2$ supersymmetric free particle. However, in all such models, the ${\mathcal N} = 2$ SUSY symmetries
are nilpotent of order two (see, e.g., Refs. [14-18] of our present paper). 

In the main body of our text, we have established that the conserved 
and nilpotent  (anti-) BRST charges  are found to be absolutely anticommuting, too, within the framework of ACSA to BRST formalism where
{\it only} the (anti-)chiral super expansions are taken into account. In our present Appendix, we show that the
application of the {\it very same} ACSA to one of the simplest  ${\mathcal N} = 2$ SUSY QM models (e.g. a free particle) does {\it not} lead to 
the absolute anticommutativity of the ${\mathcal N} = 2$ SUSY conserved and nilpotent charges. Thus, the central
objective of our present Appemdix is to establish the {\it novelty} of the observation of absolute anticommutativity 
property, associated with the (anti-)BRST charges, within the framework of ACSA to BRST formalism. Towards this goal 
in mind, let us begin with the Lagrangian  of a ${\mathcal N} = 2$ SUSY free particle (see, e.g. [16] for details)
\begin{eqnarray}
L_0 = \frac{1}{2}{\dot x}^2 + i \,\bar\psi\,{\dot\psi},
\end{eqnarray}
where $x(t)$ is a bosonic variable and ($\bar\psi(t)\,{\psi (t)}$) are the ${\mathcal N} = 2$  fermionic (i.e. ${\psi}^2 = {\bar\psi}^2 =0, \, 
\psi\,\bar\psi + \bar\psi\,\psi =0 $) SUSY partners. For the sake of breviety, we have taken the mass of the particle 
to be one (i.e. $m = 1$) and we have taken ${\dot x} = dx/dt, {\dot\psi} = d\psi/dt$ as the generalized ``velocities" for the 
${\mathcal N} = 2$ SUSY particle where the ``time" evolution parameter is $t$.

The above Lagrangian respects the following  continuous and nilpotent (${s_1}^2 = {s_2}^2 = 0$)  
 infinitesimal ${\mathcal N} = 2$ SUSY transformations 
\begin{eqnarray}
&&s_1 \, x = i\, \psi, \qquad\qquad s_1 \, \psi = 0, \qquad\qquad s_1 \, \bar\psi = - \,{\dot x}, \nonumber\\
&&s_2 \, x = i\, \bar\psi, \qquad\qquad s_2 \, \bar\psi = 0, \qquad\qquad s_2 \, \psi = -\, {\dot x},
\end{eqnarray}
because we observe that the following are  true, namely; \
\begin{eqnarray}
s_1\, L_0  = 0, \qquad \qquad s_2 \, L_0  = \frac{d}{dt}\,(i\, \dot x\, \bar\psi), 
\end{eqnarray}
which render the action integral $S = \int dt\, L_0$ invariant for the physical variables which vanish off at 
$t = \pm \, \infty$. It should be noted that the SUSY transformations are  on-shell ($\dot{\psi} = 0, \,\dot{\bar\psi} = 0$)
nilpotent  (${s_1}^2 = {s_2}^2 = 0$) of order two and they do {\it not} absolutely anticommute (i.e. $\{s_1, \, s_2\}\, \ne 0$).
According to Noether's theorem, it is straightforward to note that the following ${\mathcal N} = 2$  conserved 
($\dot Q = {\dot{\bar Q}} = 0$) and nilpotent ($Q^2 = {\bar Q}^2 = 0$)   SUSY charges
\begin{eqnarray}
Q = i\, \dot x\, \psi \equiv i\, p\, \psi,\qquad \qquad \qquad \bar Q = i\, \dot x\, \bar\psi \equiv i\, p\, \bar\psi, 
\end{eqnarray}
are the generators of the ${\mathcal N} = 2$ SUSY transformations (40) where $p = \dot x$ is the momentum 
corresponding to the bosonic variable $x(t)$.

We apply now the ACSA to derive the continuous and nilpotent symmetry transformations $s_1$. Towards this goal in 
mind, we generalize the variables ($ x(t),\, \psi(t), \, \bar\psi(t)$) of the Lagrangian (39) onto a (1, 1)-dimensional
{\it anti-chiral} super-submanifold 
(of the general (1, 2)-dimensional manifold) as [16]
\begin{eqnarray}
x(t) \longrightarrow  X(t, \bar\theta) &=& x(t) + \bar\theta f(t), \nonumber\\
\psi(t) \longrightarrow  \Psi(t, \bar\theta) &=& \psi (t) + i\, \bar\theta \, b_1(t), \nonumber\\
\bar\psi(t) \longrightarrow  \bar\Psi(t, \bar\theta) &=& \psi (t) + i\, \bar\theta \, b_2(t),
\end{eqnarray}
where $(b_1 (t), b_2 (t))$ are the {\it bosonic} secondary variables and $f(t)$ is a {\it fermionic} secondary 
variable due to the fermionic (${\bar\theta}^2 = 0$) nature of the Grassmannian variable $\bar\theta$ which 
parametrizes the (1, 1)-dimensional {\it anti-chiral} super-submanifold (along with the bosonic evolution parameter $t$). 
According to the basic tenets of ACSA, the following interesting and useful ${\mathcal N} = 2$ SUSY invariant quantities
(due to the SUSY transformations $s_1$), namely;
\begin{eqnarray}
s_1 \psi = 0, \quad\quad s_1 (x\,\psi) = o,  \quad\quad  s_1 (\dot x\, \dot\psi) = 0, 
\quad\quad  s_1 \Big(\frac{1}{2}{\dot x}^2 + i\, \bar\psi \, \dot\psi\Big) = 0,
\end{eqnarray}
should be independent of the Grassmannian variable $\bar\theta$ when they
are generalized onto the (1, 1)-dimensional {\it anti-chiral} super-submanifold:
\begin{eqnarray}
&& \Psi(t, \bar\theta) =  \psi(t), \quad X(t, \bar\theta) \, \Psi(t, \bar\theta) = x(t)\,\psi(t),
\quad {\dot X}(t, \bar\theta) \, \dot\Psi(t, \bar\theta) = {\dot x}(t)\, \dot\psi(t),\nonumber\\
&&\Big(\frac{1}{2}{\dot X}^2 (t, \bar\theta) + i\, \bar\Psi (t, \bar\theta) \, \dot\Psi (t, \bar\theta)\Big)
= \Big[\frac{1}{2}({\dot x (t)})^2  + i\, \bar\psi (t) \, \dot\psi (t)\Big].
\end{eqnarray}
The above SUSY invariant restrictions lead to the determination of the secondary variables as: 
$b_1 (t) = 0, f(t) = i\, \psi (t), b_2 (t) = i\, \dot x$ (see, e.g. [16] for details). The substitution of these
values into the super-subexpansions (43) leads to the following
\begin{eqnarray}
X^{(1)}(t, \bar\theta) &=& x(t) + \bar\theta \, \Big(i \psi (t)\Big) \equiv x(t) + \bar\theta \, \Big(s_1\, x(t)\Big), \nonumber\\
\Psi^{(1)}(t, \bar\theta) &=& \psi (t) + i\, \bar\theta \, \Big(0\Big) \equiv \psi (t) + i\, \bar\theta \, \Big(s_1\, \psi (t)\Big), \nonumber\\
{\bar\Psi}^{(1)}(t, \bar\theta) &=& \bar\psi (t) + i\, \bar\theta \, \Big(- \dot x\Big) \equiv \bar\psi (t) +
 i\, \bar\theta \, \Big(s_1\, \bar\psi (t)\Big),
\end{eqnarray}
where the superscript (1) on the anti-chiral supervariables denotes the super expansions of {\it such} supervariables which lead to 
the derivation of the SUSY transformations $s_1$ as the coefficient of $\bar\theta$. In other words, the above
supervariables (46) have been derived after the application of the SUSY invariant restrictions (45) which lead to the
derivation of the secondary variables in terms of basic variables. We also note that we have obtained a mapping:
$s_1 \longleftrightarrow \partial_{\bar\theta}$ which establishes a deep connection between $s_1$ and $\partial_{\bar\theta}$.

Against the backdrop of the above discussions, we now derive the SUSY transformations $s_2$. For this, we generalize 
the variables $x(t), \psi(t), \bar\psi(t) $ onto the (1, 1)-dimensional {\it chiral} super-submanifold as [16]
\begin{eqnarray}
x(t) \longrightarrow  X(t, \theta) &=& x(t) + \theta \bar f(t), \nonumber\\
\psi(t) \longrightarrow  \Psi(t, \theta) &=& \psi (t) + i\, \theta \, \bar b_1(t), \nonumber\\
\bar\psi(t) \longrightarrow  \bar\Psi(t, \theta) &=& \psi (t) + i\, \theta \, \bar b_2(t),
\end{eqnarray}
where the secondary variables ($\bar b_1(t), \bar b_2(t)$) are bosonic and $\bar f(t)$ is fermionic in 
nature because of the fermionic (${\theta}^2 = 0$) nature of the Grassmannian variable ${\theta}$ which 
characterizes the {\it chiral} super-submanifold (along with the bosonic evolution operator $t$). 
The following SUSY invariant quantities under the SUSY transformation $s_2$, namely; 
\begin{eqnarray}
s_1 \bar\psi = 0, \quad\quad s_1 (x\,\bar\psi) = o, \quad\quad s_1 (\dot x\, \dot{\bar\psi}) = 0, 
\quad\quad s_1 \Big(\frac{1}{2}{\dot x}^2 + i\, \dot{\bar\psi} \, {\bar\psi}\Big) = 0,
\end{eqnarray}
lead to the following restrictions on the {\it chiral} supervariables:  
\begin{eqnarray}
&&\bar\Psi(t, \theta) =  \psi(t), \qquad X(t,\theta) \, \bar\Psi(t, \theta) = x(t)\,\bar\psi(t),
\qquad {\dot X}(t, \theta) \, \dot{\bar\Psi}(t, \theta) = {\dot x}(t)\, \dot{\bar\psi}(t),\nonumber\\
&&\Big(\frac{1}{2}{\dot X}^2 (t, \theta) - i\, \dot{\bar\Psi} (t, \theta)\, \Psi (t, \theta)\Big)
= \Big[\frac{1}{2}({\dot x (t)})^2 - i\, \dot{\bar\psi} (t) \, \psi (t)\Big].
\end{eqnarray}
These restrictions lead to the derivation of the secondary variables in terms of the basic variables as: 
${\bar b}_2 (t) = 0, \bar f (t) = i\, \bar\psi (t), {\bar b}_1 (t) = i\, \dot x (t)$. 
As a consequence, we have the following supervariable expansions (after substitutions of the
secondary variables), namely; 
\begin{eqnarray}
X^{(2)}(t, \theta) &=& x(t) + \theta \, \Big(i \bar\psi (t)\Big) \equiv x(t) + \theta \, \Big(s_2\, x(t)\Big), \nonumber\\
\Psi^{(2)}(t, \theta) &=& \psi (t) + i\, \theta \, \Big(- \dot x\Big) \equiv \psi (t) + i\, \theta \, \Big(s_2\, \psi (t)\Big), \nonumber\\
{\bar\Psi}^{(2)}(t, \theta) &=& \bar\psi (t) + i\, \theta \, \Big(0\Big) \equiv \psi (t) + i\, \theta \, \Big(s_2\, \bar\psi (t)\Big),
\end{eqnarray}
where the superscript (2) on the {\it chiral} supervariables denote the expansions of the 
{\it latter} after the application of (49) and which lead to the derivation of $s_2$ as the
coefficients of $\theta$. Hence, we have a mapping: $s_2 \longleftrightarrow \partial_{\theta}$ which establishes 
an intimate relationship between $s_2$ and $\partial_{\theta}$ that leads to their nilpotency 
(${s_2}^2 = 0, \, {\partial_{\theta}}^2 = 0$) property.

We now concentrate on the discussion about the ${\mathcal N} = 2$ SUSY charges $Q$ and $\bar Q$ (cf. Eq. (42)).
These are conserved ($\dot Q =  \dot{\bar Q}  =  0$) as it can be checked by the application of the EL-EOM: $\dot p = \ddot x = 0, 
\dot\psi = \dot{\bar\psi} = 0$. They are nilpotent (i.e. ${Q}^2 =  {\bar Q}^2 = 0$)  of order two 
as it can be verified that we can express them in {\it two} different ways: 
\begin{eqnarray}
&&Q = s_1 (- \, i\, \bar\psi (t)\, \psi (t)), \qquad\qquad Q = s_1 (\dot x (t)\, x(t)),\nonumber\\ 
&&\bar Q = s_2 (i\, \bar\psi (t)\, \psi (t)), \quad\qquad\qquad  Q = s_2 (\dot x (t)\, x(t)),
\end{eqnarray}
provided we use the on-shell conditions: $\dot\psi = 0$ and $\dot{\bar\psi} = 0$. We lay emphasis on the
fact that $Q$ can {\it not} be expressed in terms of $s_2$ and $\bar Q$, in no way, could be expressed in terms of $s_1$.
Using the basic principle behind the continuous symmetries and their generators, it is straightforward to 
check that the following are true, namely; 
\begin{eqnarray}
s_1 \,Q = -\,i\, \{Q, Q\} = 0 \quad \Longleftrightarrow \quad Q^2 = 0 \quad \Longleftrightarrow \quad {s_1}^2 = 0, \nonumber\\
s_2 \,Q = -\,i\, \{{\bar Q}, {\bar Q}\} = 0 \quad \Longleftrightarrow \quad {\bar Q}^2 = 0 \quad \Longleftrightarrow \quad {s_2}^2 = 0. 
\end{eqnarray}
The above equations {\it clearly} demonstrate that the nilpotency (${s_1}^2 = 0, Q^2 = 0$) property 
of the ${\mathcal N} = 2$ SUSY transformations $s_1$ and its generator $Q$ are intertwined in a meaningful
and beautiful manner. The same statement could be repeated for $s_2$ and $\bar Q$ as well. 
We can capture the observations in (51) and (52),
 within the framework of ACSA to our present 
${\mathcal N} = 2$ SUSY QM model of a free particle, as: 
\begin{eqnarray}
Q &=& \frac{\partial}{\partial \bar\theta} \Big[-\,i\, \bar\Psi^{(1)}(t, \bar\theta)\,\Psi^{(1)}(t, \bar\theta)\Big] 
\equiv \int d\bar\theta \;\Big[-\,i\, \bar\Psi^{(1)}(t, \bar\theta)\,\Psi^{(1)}(t, \bar\theta)\Big],\nonumber\\
Q &=& \frac{\partial}{\partial \bar\theta} \Big[{\dot X}^{(1)}(t, \bar\theta)\,X^{(1)}(t, \bar\theta)\Big] 
\equiv \int d\bar\theta \;\Big[{\dot X}^{(1)}(t, \bar\theta)\,X^{(1)}(t, \bar\theta)\Big],\nonumber\\
{\bar Q} &=& \frac{\partial}{\partial \theta} \Big[i\,\Psi^{(2)}(t, \theta)\,\Psi^{(2)}(t, \theta)\Big] 
\equiv \int d\theta \;\Big[i\, \bar\Psi^{(2)}(t, \theta)\,\Psi^{(2)}(t, \theta)\Big],\nonumber\\
{\bar Q} &=& \frac{\partial}{\partial \theta} \Big[{\dot X}^{(2)}(t, \theta)\,X^{(2)}(t, \theta)\Big]
\equiv \int d\theta \;\Big[{\dot X}^{(2)}(t, \theta)\,X^{(2)}(t, \theta)\Big].
\end{eqnarray}
It s very evident, from the above, that we have the following:
\begin{eqnarray}
\partial_{\bar\theta}\, Q = 0 \quad \Longleftrightarrow \quad {\partial_{\bar\theta}}^2 = 0 \quad  \Longleftrightarrow \quad {s_1}^2 = 0 
\quad \Longleftrightarrow \quad Q^2 = 0, \nonumber\\
\partial_{\theta}\, {\bar Q} = 0 \quad \Longleftrightarrow \quad {\partial_{\theta}}^2 = 0 \quad \Longleftrightarrow \quad {s_2}^2 = 0
 \quad \Longleftrightarrow \quad {\bar Q}^2 = 0. 
\end{eqnarray}
In other words, we note that the nilpotency properties (${s_1}^2 =  Q^2 = 0,\, {s_2}^2 = {\bar Q}^2 = 0$) of the 
${\mathcal N} = 2$ SUSY transformations as well as their corresponding charges and the nilpotency (${\partial_{\theta}}^2 = 
 {\partial_{\bar\theta}}^2 = 0$)  of the translational generators ($\partial_{\bar\theta}, \,
 \partial_{\theta}$) along the Grassmannian directions of the {\it (anti-)chiral} super-submanifolds
 (of the general (1, 2)-dimensional supermanifold) are found to blend together in a beautiful and intimate fashion.

We end this Appendix with the following crucial and decisive remark. It has been amply emphasized that
the ${\mathcal N} = 2$ SUSY charges ($Q$ and $\bar Q$) can {\it not} be expressed in terms of $s_1$ and $s_2$ {\it together}. In
other words, the charge $Q$ can be expressed  {\it only} as the {\it exact} form w.r.t. $s_1$ and
there is {\it no} way to express it as the {\it exact} form w.r.t. $s_2$. Similar statement can be made for ${\bar Q}$, 
too. On the contrary, we have shown in Sec. 6 that the (anti-)BRST charges can be expressed as the {\it exact} form
w.r.t. BRST and anti-BRST symmetry transformations. Similar statement can be made for the (anti-)co-BRST charges as well. 
Translated in the language of ACSA, both the (anti-)BRST and (anti-)co-BRST charges can be expressed as the derivatives
w.r.t. {\it both} the translational generators ($\partial_{\bar\theta}, \, \partial_{\theta}$) {\it but} the {\it same}
is {\it not} true for the ${\mathcal N} = 2$ SUSY charges. Thus, the observation of the absolute anticommutativity
for the (anti-)BRST and (anti-)co-BRST charges is a {\it novel} and surprising observation ({\it vis-{\`a}-vis} the ${\mathcal N} = 2$ SUSY
conserved and nilpotent charges) within the framework of ACSA.

\end{document}